\documentstyle[psfig,12pt]{article}
\def\lsim{\:\raisebox{-0.5ex}{$\stackrel{\textstyle<}{\sim}$}\:}

\def\VEV#1{\left\langle #1\right\rangle}
\def\be{\begin{equation}}       
\def\ee{\end{equation}}
\def\bear{\be\begin{array}}      
\def\eear{\end{array}\ee}
\def\bea{\begin{eqnarray}}
\def\eea{\end{eqnarray}}

\def\21{$SU(2) \ot U(1)$}
\def\ot{\otimes}

\def\etal{{\it et al.}}

\def\quarter{{\textstyle{1 \over 4}}}

\def\bold#1{\setbox0=\hbox{$#1$}
     \kern-.025em\copy0\kern-\wd0
     \kern.05em\copy0\kern-\wd0
     \kern-.025em\raise.0433em\box0 }
\begin{document}
\begin{titlepage}
\begin{flushright}
FTUV/98-9\\
IFIC/98-9\\
FISIST/1-98/CFIF\\
hep-ph/9801391 \\
January 1998
\end{flushright}
\vspace*{5mm}
\begin{center} 
{\Large \bf Gauge and Yukawa Unification with Broken R--Parity
}\\[15mm]
{\large Marco A. D\'\i az${}^1$, J. Ferrandis${}^1$, 
Jorge C. Rom\~ao${}^2$ 
and 
{\large Jos\'e W. F. Valle${}^1$}\\
\hspace{3cm}\\
{\small ${}^1$Departamento de F\'\i sica Te\'orica, IFIC-CSIC, 
Universidad de Valencia}\\ 
{\small Burjassot, Valencia 46100, Spain}
\hspace{3cm}\\
\vskip .3cm
{\small ${}^2$Departamento de F\'\i sica, Instituto Superior T\'ecnico}\\ 
{\small A. Rovisco Pais, 1096 Lisboa Codex, Portugal}} 
\end{center}
\vspace{5mm}
\begin{abstract}

We study gauge and Yukawa coupling unification in the simplest
extension of the Minimal Supersymmetric Standard Model (MSSM) which
incorporates R--Parity violation through a bilinear superpotential
term. Contrary to what happens in the MSSM, we show that bottom-tau
unification at the scale $M_{GUT}$ where the gauge couplings unify can
be achieved for any value of $\tan\beta$ by choosing appropriately the
sneutrino vacuum expectation value. In addition, we show that
bottom-tau-top unification occurs in a slightly wider $\tan\beta$
range than in the MSSM.

\end{abstract}

\end{titlepage}

\setcounter{page}{1}

The Standard Model (SM) of particle physics is very successful in
describing the interactions of the elementary particles, except
possibly neutrinos. Although it is regarded as a good low-energy
effective theory, the SM has many unanswered questions and theoretical
problems. Its gauge symmetry group is the direct product of three
groups $SU(3)\times SU(2)\times U(1)$ and the corresponding gauge
couplings are unrelated. It does not explain the three family
structure of quarks and leptons, and their masses are fixed by
arbitrary Yukawa couplings, with neutrinos being prevented from having
mass. The Higgs sector, responsible for the electroweak symmetry
breaking and for the fermion masses, has not been verified
experimentally and the Higgs mass is unstable under radiative
corrections. As a result, say, the hierarchy between the electroweak
scale and the Planck scale is not understood.

In supersymmetry (SUSY) \cite{SUSY} the Higgs mass is stabilized
under radiative corrections because the loops containing standard
particles is partially cancelled by the contributions from loops
containing supersymmetric particles. If we add to the Minimal
Supersymmetric Standard Model (MSSM) \cite{MSSM} the notion of Grand
Unified Theory (GUT), then we find that the three gauge couplings
unify at a certain scale $M_{GUT}$ \cite{GUT}. Indeed, measurements of
the gauge couplings at the CERN $e^+e^-$ collider LEP and neutral
current data \cite{PDG} are in good agreement with the MSSM--GUT with
the SUSY scale $M_{SUSY}\lsim 1$ TeV \cite{gaugeUnif}. In addition, the
unification scale in SUSY--GUT is high enough to predict a proton
decay rate slower than present experimental limits, as opposed to the
non--SUSY GUTs, where the proton decays too fast.

Besides achieving gauge coupling unification \cite{gaugUnifRecent},
GUT theories also reduce the number of free parameters in the Yukawa
sector.  For example, in $SU(5)$ models, the bottom quark and the tau
lepton Yukawa couplings are equal at the unification scale, and the
predicted ratio $m_b/m_{\tau}$ at the weak scale agrees with
experiments. Furthermore, a relation between the top quark mass and
$\tan\beta$, the ratio between the vacuum expectation values of the
two Higgs doublets is predicted. Two solutions are possible,
characterized by low and high values of $\tan\beta$ \cite{YukUnif}.
In models with larger groups, such as $SO(10)$ and $E_6$, both the top
and bottom Yukawa couplings are unified with the tau Yukawa at the
unification scale \cite{YukUniThree}. In this case, only the large
$\tan\beta$ solution survives.

Recent global fits of low energy data to minimal supersymmetry
\cite{deBoerJeru} show that it is hard to reconcile these constraints
with the large $\tan\beta$ solution. Specially important are the
measurements of the $B(b\rightarrow s\gamma)$ decay rate and the bound
on the lightest Higgs mass. In addition, the low $\tan\beta$ solution
with $\mu<0$ is also disfavoured. In this letter, we show that the
minimal extension of the MSSM--GUT \cite{epsrad} in which R--Parity
violation is introduced via a bilinear term in the MSSM superpotential
\cite{e3others,chaHiggsEps}, allows $b-\tau$ Yukawa unification for
any value of $\tan\beta=v_u/v_d$ and satisfying perturbativity of the
couplings.  We also analize the $t-b-\tau$ Yukawa unification and find
that it is easier to achieve than in the MSSM, occurring in a slightly
wider high $\tan\beta$ region.

For simplicity, we consider only the third generation of quarks and
leptons.  In this way, the superpotential is given by
\begin{equation} 
W=
 h_t \widehat Q_3 \widehat U_3\widehat H_u
+h_b \widehat Q_3 \widehat D_3\widehat H_d
+h_{\tau}\widehat L_3 \widehat R_3\widehat H_d
+\mu\widehat H_u \widehat H_d
+\epsilon_3\widehat L_3 \widehat H_u 
\label{eq:Wsuppot}
\end{equation}
where the first four terms correspond to the MSSM and the last one is
the bilinear term which violates R--parity. This superpotential is 
motivated by models of spontaneous breaking of R--Parity \cite{MV_RIV}. 
Here, R--Parity and lepton number are violated explicitly by the 
$\epsilon_3$ term.

It is clear from eq.~(\ref{eq:Wsuppot}) that the scalar potential contains
terms which induce a non--zero vacuum expectation value (VEV) of the
tau sneutrino $\VEV{\tilde\nu_{\tau}}=v_3/\sqrt{2}$. It contributes to
the $W$ mass according to $m_W^2=\quarter g^2(v_d^2+v_u^2+v_3^2)$,
where $v_d/\sqrt{2}$ and $v_u/\sqrt{2}$ are the VEVs of the two Higgs
doublets $H_d$ and $H_u$ respectively. The R--Parity violating
parameters $\epsilon_3$ and $v_3$ violate tau--lepton number, inducing
a non-zero $\nu_{\tau}$ mass 
$m_{\nu_{\tau}}\propto (\mu v_3+\epsilon_3v_d)^2$, which arises due to 
mixing between the weak eigenstate $\nu_{\tau}$ and the neutralinos. The 
latest $\nu_{\tau}$ mass limit from ALEPH is $m_{\nu_{\tau}} \lsim 16$ MeV. 
The $\nu_e$ and $\nu_{\mu}$ remain massless in first approximation.  They 
acquire typically smaller masses from supersymmetric loops.  As already
mentioned, in what follows we consider only the third generation of
quarks and leptons.

It is important to note that the $\epsilon$--term in eq.~(\ref{eq:Wsuppot})
is a physical parameter and cannot be eliminated by a redefinition of
the superfields $\widehat H_d$ and $\widehat L_3$ \cite{HallSuzuki}.
The reason is that, after the rotation, bilinear terms which induce a
tau sneutrino VEV are re--introduced in the soft scalar sector
\cite{Basis}. Moreover, in contrast to many prejudices \cite{Banks}, 
we wish to stress that the R--Parity violating parameters $v_3$ and
$\epsilon_3$ need not be small. In models with universality of soft
supersymmetry breaking mass parameters \cite{Basis} $m_{\nu_{\tau}}$
is naturally small because it arises from a seesaw mechanism in which
the the {\sl effective} mixing arises only radiatively, and can easily
lie in the eV range \cite{epsrad}.

R--Parity violation also implies that the charginos mix with the tau
lepton, through a mass matrix is given by
\begin{equation} 
{\bold M_C}=\left[\matrix{ 
M & {\textstyle{1\over{\sqrt{2}}}}gv_u & 0 \cr 
{\textstyle{1\over{\sqrt{2}}}}gv_d & \mu &  
-{\textstyle{1\over{\sqrt{2}}}}h_{\tau}v_3 \cr 
{\textstyle{1\over{\sqrt{2}}}}gv_3 & -\epsilon_3 & 
{\textstyle{1\over{\sqrt{2}}}}h_{\tau}v_d}\right] 
\label{eq:ChaM6x6} 
\end{equation} 
with $h_{\tau}$ being the tau Yukawa coupling.  Imposing that one of
the eigenvalues reproduces the observed tau mass $m_{\tau}$, the tau
Yukawa coupling can be solved exactly as \cite{chaHiggsEps}
\begin{equation}
h_{\tau}^2={{2m_{\tau}^2}\over{v_d}}\left[
{{1+\delta_1}\over{1+\delta_2}}
\right]
\label{appYukTau}
\end{equation}
where the $\delta_i\,$, $i=1,2$, depend on $m_{\tau}$, on the SUSY
parameters $M,\mu,\tan\beta$ and on the R-parity violating parameters
$\epsilon_3$ and $v_3$.  They can be found in ref. \cite{chaHiggsEps}
and can easily be shown to vanish in the MSSM limit $\epsilon_3 \to 0$
and $v_3 \to 0$. On the other hand, the bottom and top Yukawa
couplings are related to the bottom and top masses according to
\be
m_t = h_t \frac{v}{\sqrt2} \sin \beta \sin \theta\,, \: \: \: \: \:
m_b = h_b \frac{v}{\sqrt2} \cos \beta \sin \theta 
\ee
where we use spherical coordinates for the VEVs, defining
$v=2m_W/g$, $\tan\beta=v_u/v_d$, and $\cos\theta=v_3/v$. 

We now turn to the study of the renormalization group evolution of the
various relevant parameters of the model such as the gauge and Yukawa
couplings, the SM quartic Higgs coupling and the third generation
fermion masses.  In our approach we divide the evolution into three
ranges: (i) from $M_{SUSY}$ to $M_{GUT}$, where we use the two-loop
RGEs of our model, (ii) from $m_t$ to $M_{SUSY}$, where we use the
two-loop SM RGEs including the quartic Higgs coupling and (iii) from
$M_{Z}$ to $m_t$ we use running fermion masses and gauge couplings.

Using a top-bottom approach we randomly vary the unification scale
$M_{GUT}$ and the unified coupling $\alpha_{GUT}$ looking for
solutions compatible with the low energy data \cite{LEPinternal}
$\alpha^{-1}_{em}(m_Z) = 128.896 \pm0.090$, $\sin^2\theta_w(m_Z) =
0.2322 \pm 0.0010$, and $\alpha_s(m_Z)=0.118 \pm 0.003$.  We use the
approximation of a common decoupling scale $M_{SUSY} \lsim 1$ TeV for
all the supersymmetric particles.  The solutions we find are
concentrated in a region of the $M_{GUT}-\alpha_{GUT}$ plane. For the
simpler case where the SUSY scale coincides with the top mass,
$M_{SUSY}=m_t$, this region is centered at the point $M_{GUT} \approx
2.3 \times10^{16}$ GeV and ${\alpha_{GUT}}^{-1} \approx 24.5$, which
we adopt from now on.

Next, we study the unification of Yukawa couplings using two-loop
RGEs. We take $m_W = 80.41 \pm 0.09$ GeV, $m_{\tau}=1777.0 \pm 0.3$ 
MeV, and $m_b(m_b)$ = 4.1 to 4.5 GeV \cite{LEPinternal}. 
We calculate the running masses
$m_{\tau}(m_t)=\eta_{\tau}^{-1}m_{\tau}(m_{\tau})$ and
$m_b(m_t)=\eta_b^{-1}m_b(m_b)$, where $\eta_{\tau}$ and $\eta_b$
include three--loop order QCD and one--loop order QED \cite{alf3}. At
the scale $Q=m_t$ we keep as a free parameter the running top quark
mass $m_t(m_t)$ and vary randomly the SM quartic Higgs coupling
$\lambda$. Using SM RGEs we evolve the gauge, Yukawa, and Higgs
couplings from $Q=m_t$ up to $Q=M_{SUSY}$. The initial conditions for
the SM Yukawa couplings are $\lambda_i^2(m_t)=2m_i^2(m_t)/v^2$, with
$i=t,b,\tau$ and $v=246.2$ GeV.

At the scale $Q=M_{SUSY}$, below which all SUSY particles are
decoupled (including the heavy Higgs bosons) we impose the following
boundary conditions for the quark Yukawa couplings
\begin{eqnarray}
\lambda_t(M_{SUSY}^-)=h_t (M_{SUSY}^+) \sin\beta\sin\theta 
\nonumber\\
\lambda_b(M_{SUSY}^-)=h_b (M_{SUSY}^+) \cos\beta\sin\theta 
\label{BounQYuk}
\end{eqnarray}
where $h_i$ denote the Yukawa couplings of our model and $\lambda_i$
those of the SM. Due to its mixing with charginos, the boundary
condition for the tau Yukawa coupling is slightly more complicated:
\begin{equation}
\lambda_{\tau}(M_{SUSY}^-)=h_{\tau} (M_{SUSY}^+) \cos\beta\sin\theta
\sqrt{{1+\delta_2}\over{1+\delta_1}}\,
\label{BounTauYuk}
\end{equation}
Finally, the boundary condition for the quartic Higgs coupling is
given by 
\begin{equation}
\lambda(M_{SUSY}^-) = 
\frac{1}{4}
\Big[(g^2(M_{SUSY}^+)+g'^2(M_{SUSY}^+) \Big] (\cos2\beta\sin^2\theta+
\cos^2\theta)^2
\label{BounHiggsCoup}
\end{equation}
The MSSM limit is obtained setting $\theta \to \pi/2$ i.e. $v_3=0$.

At the scale $Q=M_{SUSY}$ we vary randomly the SUSY parameters $M$,
$\mu$ and $\tan\beta$, as well as the R--Parity violating parameter
$\epsilon_3$. The parameter $v_3=v\cos\theta$ is calculated from
eq.~(\ref{BounHiggsCoup}). Since $\lambda$ (or equivalently the SM Higgs 
mass $m_H^2=2\lambda v^2$) is varied randomly, in practice we also scan
over $\theta$. This way, we consider all possible initial conditions
for the RGEs at $Q=M_{SUSY}$, and evolve them up to the unification
scale $Q=M_{GUT}$.  The solutions that satisfy $b-\tau$ unification
are kept.

\begin{figure}
\centerline{\protect\hbox{ 
\psfig{file=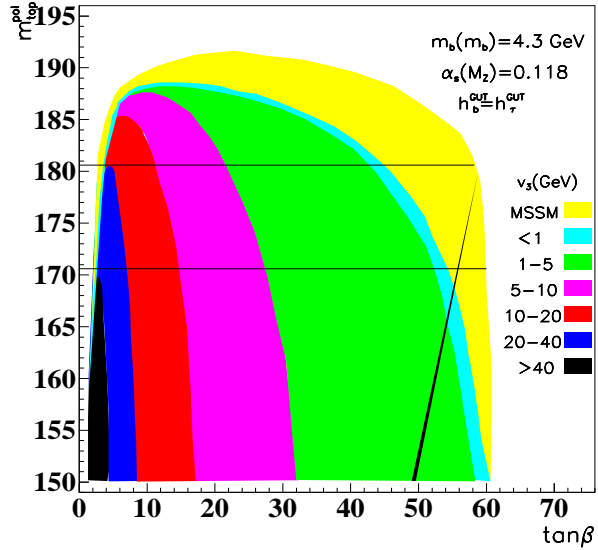,height=8.5truecm,width=9truecm}}}
\caption{Top quark mass as a function of $\tan\beta$ for different
values of the R--Parity violating parameter $v_3$. Bottom quark and
tau lepton Yukawa couplings are unified at $M_{GUT}$. The horizontal
lines correspond to the $1\sigma$ experimental $m_t$
determination. Points with $t-b-\tau$ unification lie in the diagonal
band at high $\tan\beta$ values. We have taken $M_{SUSY}=m_t$.}
\label{aretop}
\end{figure}
In Fig.~\ref{aretop} we illustrate our point by plotting the top quark
mass (we always use the pole mass) as a function of $\tan\beta$. For
simplicity we have taken $M_{SUSY}=m_t$ but it should be clear that a
different $M_{SUSY}$ choice would not change qualitatively our
results.  Each selected point in our scan satisfies bottom-tau
unification (to within a $1\%$) $h_b(M_{GUT})=h_{\tau}(M_{GUT})$ and
it is placed in one of the shaded regions according to the value of
$|v_3|$. The first region with $v_3=\epsilon_3=0$ corresponds to the
MSSM and sits at the top of the plot. Points with $|v_3|<1$ GeV fall
in the region just below. The subsequent regions labelled by
$1<|v_3|<5$ GeV up to $|v_3|>40$ GeV respectively are obtained when
$v_3$ gets higher.  They are narrower in $\tan \beta$. Note that
points with smaller $v_3$ values, say $1<|v_3|<5$ GeV fall in the
region labelled as such, as well as in all previous regions, but not
in the subsequent ones. This overlapping with the previous regions
decreases as we increase $|v_3|$ in such a way that points with
$|v_3|>40$ GeV fall almost exclusively in the last region.

The two horizontal lines correspond to the top quark mass within a
$1\sigma$ error. In the MSSM limit we can see the two solutions
compatible with the experimental value of the top quark mass, one with
$\tan\beta\approx1$ and the other with $\tan\beta\approx$ 55-60.  It
is clear from the figure that by selecting appropriately the value of
$|v_3|$ we can find $b-\tau$ unification for any $\tan\beta$ value
within the perturbative region $1\lsim\tan\beta\lsim62$ of the Yukawa
couplings. For $|v_3| \lsim 20$ GeV one has, as in the MSSM, two
disconnected solutions for $b-\tau$ unification, one with
$\tan\beta\approx1$, and a large $\tan\beta$ range which, for
intermediate $v_3$ can be quite broad. Note that for $20<|v_3|<40$ GeV
only the $\tan\beta$ range from 3 to 8 or so is consistent with the 1
$\sigma$ top mass measurement, for the chosen $\alpha_s$ and
$m_b(m_b)$ values. Similarly, the $|v_3|$ range above 40 GeV would be
ruled out. Note that our results do not depend qualitatively on the
definition chosen for $\tan\beta$. For example, if we define
$\tan\beta$ in the way which is natural in the basis where the
$\epsilon_3$--term disappears from the superpotential,
$\tan\beta'\equiv v_u/\sqrt{v_d^2+v_3^2}$ we also can find $b-\tau$
unification for any $\tan\beta'$ value.

We now turn to the discussion of the uncertainties of the unification
program in this model. The general trend follows closely that of the
MSSM. The dependence of our results on the strong coupling constant
and the bottom mass running is totally analogous to what happens in
the MSSM.  Indeed, we have studied the effect of varying $\alpha_s$ in
Fig.~\ref{aretop} and found that the upper bound on $\tan\beta$, which
is $\tan\beta\lsim 61$ for $\alpha_s=0.118$, increases with $\alpha_s$
and becomes $\tan\beta\lsim 63$ (59) for $\alpha_s=0.122$ (0.114). On
the other hand the MSSM region is narrower if $\alpha_s$ increases,
specially at high $\tan\beta$ values. We have verified that the same
trend extends to the regions with large $v_3$. Finally, we mention
that the top mass value for which unification is achieved for any
$\tan\beta$ value within the perturbative region increases with
$\alpha_s$, as in the MSSM.  Turning to the dependence on $m_b$, the
behaviour is the opposite one. In Fig.~\ref{aretop} we have taken
$m_b(m_b)=4.3$ GeV. As before the value of $\tan\beta$ is bounded from
above by $\tan\beta\lsim 61$ due to the perturbativity condition of
the bottom quark Yukawa coupling. If we consider $m_b(m_b)=4.1$ (4.5)
GeV then the upper bound of this parameter is given by $\tan\beta\lsim
64$ (58). In addition, the MSSM region is narrower (wider) at high
$\tan\beta$ compared with the $m_b(m_b)=4.3$ GeV case shown in
Fig.~\ref{aretop}.

Finally we have studied the possibility of bottom-tau-top unification
in our model. The diagonal line at high $\tan\beta$ values corresponds
to points where $t-b-\tau$ unification is achieved. Since the region
with $|v_3|<5$ GeV overlaps with the MSSM region, it follows that
$t-b-\tau$ unification is possible in this model for values of $|v_3|$
up to about 5 GeV, instead of 50 GeV or so, which holds in the case of
bottom-tau unification. Within the MSSM, $t-b-\tau$ unification is
achieved in the range $56 \lsim \tan \beta \lsim 59$ with $m_t$
completely inside the $1\sigma$ region. In this case, bilinear
R--Parity violation does not enlarge the allowed $\tan\beta$ region.
However, at the $2\sigma$ level our model allows $t-b-\tau$
unification for $54\lsim\tan\beta\lsim59$. In addition, we have
checked that the region with $t-b-\tau$ unification in the MSSM case
shrinks if $\alpha_s$ is increased. The space left out by the MSSM is
taken over by the regions with $|v_3|<5$ GeV so that, for large
$\alpha_s$, $t-b-\tau$ unification occurs in a wider $\tan\beta$ range
than possible in the MSSM, even in the $1\sigma$ level.

In conclusion, we have summarized \cite{new} the results of the first
systematic study of gauge and Yukawa coupling unification in a model
where we introduce bilinear R--Parity violation. The model is the
simplest alternative to the MSSM which mimics in an effective way many
of the features of models of spontaneous breaking of R--Parity.  We
showed that, in contrast to the MSSM, where bottom-tau unification is
achieved in two disconnected $\tan\beta$ regions, in our model
$b-\tau$ unification occurs for any $\tan\beta$ value, provided we
choose appropriately the value of the tau sneutrino vacuum expectation
value $v_3$. In addition, we showed that $t-b-\tau$ unification is
achieved for $|v_3|\lsim 5$ GeV at high values of $\tan\beta$ in a
slightly wider region than that of the MSSM.  Apart from the intrinsic
interest in the study of broken R--parity models, because of their
theoretical as well as phenomenological importance, our results are
relevant in connection with LEP bounds on the Higgs boson mass and
recent measurements of the $B(b\rightarrow s\gamma)$ decay rate. Taken
at face value, these disfavour the high $\tan\beta$ solution and also
the low $\tan\beta$ solution with $\mu<0$ in the MSSM as suggested in
\cite{deBoerJeru}.

\section*{Acknowledgements}

This work was supported by DGICYT under grants PB95-1077 and HP97-0039
(Accion Integrada Hispano-Portuguesa) and by the TMR network grant
ERBFMRXCT960090 of the European Union. M. A. D. was supported by a
DGICYT postdoctoral grant, J. F. was supported by a Spanish MEC FPI
fellowship.



\begin{thebibliography}{99}
\bibitem{new} 
A more comprehensive description will be presented elsewhere.
\end{thebibliography}
\end{document}